\documentclass[usenatbib]{nature_djm}
\usepackage{balance}
\usepackage{graphicx}
\usepackage{widetext}

\newcommand{\newqsolong}
  {ULAS~J112001.48+064124.3}
\newcommand{\newqso}
  {ULAS~J1120+0641}
\newcommand{\seven}
  {7.085}
\newcommand{\sevenerr}
  {7.085\mbox{\,$\pm$\,}0.003}

\newcommand{\figspectrum}
  {Fig.~1}
\newcommand{\figigm}
  {Fig.~2}
\newcommand{\figproxzone}
  {Fig.~3}
\newcommand{\figwing}
  {Fig.~4}
\newlength{\figwidth}
\newlength{\thinfigwidth}
\newlength{\tinyfigwidth}
\newlength{\widefigwidth}
\newcommand{\tq}
  {T_{\rm{q}}}
\newcommand{\rnz}
  {R_{\rm{NZ}}}
\newcommand{\rnzcorr}
  {R_{\rm{NZ,corr}}}
\newcommand{\ion}
  {{\rm{ion}}}
\newcommand{\taueff}
  {\tau_{\rm{eff}}}
\newcommand{\mabs}
  {M_{1450,{\rm{AB}}}}
\newcommand{\abs}
  {{\rm{abs}}}
\renewcommand{\micron}
  {$\mu$m}

\newcommand{\nf}
  {f_{\rm{HI}}}
\renewcommand{\sun}
  {{\rm{Sun}}}

\newcommand{\unit}[1]
  {{\mbox{\rm\,\,#1}}}

\newcommand{\lya}
  {{\rm{Ly\,\mbox{$\alpha$}}}}
\newcommand{\nv}
  {N\,{\sc{v}}}
\newcommand{\lyb}
  {Ly\,$\beta$}

\newcommand{\mgii}
  {{\rm{Mg\,{\sc{ii}}}}}
\newcommand{\siiii}
  {Si\,{\sc{iii}}]}
\newcommand{\ciii}
  {C\,{\sc{iii}}]}
\newcommand{\civ}
  {C\,{\sc{iv}}}

\newcommand{\simm}
  {\sim \!}

\newcommand{\mbh}
  {M_{\rm{BH}}}

\newcommand{\hi}
  {{\rm{H\,{\sc{i}}}}}
\newcommand{\hii}
  {{\rm{H\,{\sc{ii}}}}}

\newcommand{\ab}
  {{\rm{AB}}}

\newcommand{\J}
  {J}

\newcommand{\redshift}
  {z}

\pubdate{30 JUNE 2011}

\title[ ]
  {A luminous quasar at a redshift of \boldmath{$z = 7.085$}}

\author[ ]
  {Daniel J.\ Mortlock$^{1}$,
  Stephen J.\ Warren$^1$,
  Bram P.\ Venemans$^2$,
  Mitesh Patel$^1$,
  \newauthor
  Paul C.\ Hewett$^3$,
  Richard G.\ McMahon$^3$,
  Chris Simpson$^4$,
  Tom Theuns$^{5,6}$,
  \newauthor
  Eduardo A.\ Gonz\'{a}les-Solares$^3$,
  Andy Adamson$^7$,
  Simon Dye$^8$,
  Nigel C.\ Hambly$^9$,
  \newauthor
  Paul Hirst$^{10}$,
  Mike J.\ Irwin$^3$,
  Ernst Kuiper$^{11}$,
  Andy Lawrence$^9$
  \newauthor
  \&
  Huub J.\ A.\ R\"{o}ttgering$^{11}$}

\begin{document}

\maketitle

\newcommand{\affiliations}
{
\begin{widetext}
\small{\textsf{
\hspace*{-8mm}
$^{1}$Astrophysics Group, 
  Imperial College London, Blackett Laboratory,
  Prince Consort Road, London SW7 2AZ, UK.
$^{2}$European Southern Observatory, 2 Karl-Schwarzschild Strasse,
  85748 Garching bei M\"{u}nchen, Germany.
$^{3}$Institute of Astronomy, Madingley Road, Cambridge CB3 0HA, UK.
$^{4}$Astrophysics Research Institute, Liverpool John Moores University,
  Twelve Quays House, Egerton Wharf, Birkenhead CH41 1LD, UK.
$^{5}$Institute for Computational Cosmology, Department of Physics,
  University of Durham, South Road, Durham DH1 3LE, UK.
$^{6}$Universiteit Antwerpen, Campus Groenenborger, Groenenborgerlaan 171,
  B-2020 Antwerpen, Belgium.
$^{7}$Joint Astronomy Centre, 660 North A'oh\={o}k\={u} Place,
  Hilo, Hawaii 96720, USA.
$^{8}$School of Physics and Astronomy, University of Nottingham,  
  University Park, Nottingham NG7 2RD, UK.
$^{9}$Institute for Astronomy, SUPA (Scottish Universities Physics Alliance),
  University of Edinburgh, Royal Observatory, Blackford Hill,
  Edinburgh EH9 3HJ, UK.
$^{10}$Gemini Observatory, 670 North A'oh\={o}k\={u} Place,
  Hilo, Hawaii 96720, USA.
$^{11}$Leiden Observatory, Leiden University, PO Box 9513,
  NL-2300 RA Leiden, The Netherlands.
}
}
\end{widetext}
}

\maketitle



\noindent
{\textbf{The intergalactic 
medium was not completely reionized until 
approximately a billion years after the Big Bang,
as revealed\cite{Fan_etal:2006b} 
by observations of quasars with redshifts of less than 6.5.
It has been difficult to probe to higher redshifts, however, because 
quasars have historically been 
identified\cite{Fan_etal:2001,Fan_etal:2003,Willott_etal:2010} 
in optical surveys, which are insensitive to sources at 
redshifts exceeding 6.5.
Here we report observations of a quasar
(\newqsolong) at a redshift of \seven, which 
is 0.77 billion years after the Big Bang.
\newqso\ had a luminosity of \boldmath{$6.3\times10^{13} L_\sun$}
and hosted a black hole with a mass of \boldmath{$2 \times 10^9\, M_\sun$} 
(where \boldmath{$L_\sun$}
and $\boldmath{M_\sun}$ are the luminosity and mass of the Sun).
The measured radius of the ionized near zone around
\newqso\ was 1.9 megaparsecs,
a factor of three smaller than typical for quasars at 
redshifts between 6.0 and 6.4.
The near zone transmission profile
is consistent with a Ly\,\boldmath{$\alpha$}
damping wing\cite{Miralda-Escude:1998}, suggesting that the neutral
fraction of the intergalactic medium
in front of \newqso\ exceeded 0.1.}}



\newqso\ was first identified in the 
United Kingdom Infrared Telescope (UKIRT)
Infrared Deep Sky Survey\cite{Lawrence_etal:2007} (UKIDSS)
Eighth Data Release,
which took place on 3 September 2010.
The photometry from UKIDSS, the
Sloan Digital Sky Survey\cite{York_etal:2000} (SDSS)
and follow-up observations on UKIRT and the Liverpool Telescope
(listed in \figspectrum)
was consistent\cite{Mortlock_etal:2011a} with a 
quasar of 
redshift $\redshift \ga 6.5$.
Hence,
a spectrum was obtained 
using
the Gemini Multi-Object Spectrograph on the Gemini North Telescope
on the night beginning 27 November 2010.
The absence of significant emission blueward of a sharp break at 
$\lambda = 0.98 \unit{\micron}$
confirmed \newqso\ as a quasar
with a preliminary redshift of $\redshift = 7.08$.
Assuming a fiducial flat 
cosmological model\cite{Dunkley_etal:2009}
(that is, cosmological density parameters $\Omega_{\rm{m}} = 0.26$, 
$\Omega_{\rm{b}} = 0.024$,
$\Omega_\Lambda = 0.74$ and 
current value of the Hubble parameter
$H_0 = 72 \unit{km} \unit{s}^{-1} \unit{Mpc}^{-1}$), \newqso\ is seen as
it was 12.9 billion years (Gyr) ago, 
when the Universe was $0.77 \unit{Gyr}$ old.
While three sources have been 
spectroscopically confirmed to have even higher redshifts,
two are faint $\J_\ab \ga 26$
galaxies\cite{Lehnert_etal:2010,Vanzella_etal:2011}
and the other is a $\gamma$-ray burst
which has since faded\cite{Tanvir_etal:2009}.
Indeed, 
it has not been
possible to obtain high signal-to-noise ratio spectroscopy of 
any sources beyond the most distant quasars previously known:
CFHQS~J0210$-$0456\cite{Willott_etal:2010b}
($\redshift = 6.44$),
SDSS~1148$+$5251\cite{Fan_etal:2003}
($\redshift = 6.42$)
and
CFHQS~J2329$+$0301\cite{Willott_etal:2007}
($\redshift = 6.42$).
Follow-up measurements of \newqso\
will provide the first opportunity to explore the 
$0.1 \unit{Gyr}$ 
between $\redshift = 7.08$ and $\redshift = 6.44$,
a significant cosmological epoch about which little is currently known.


\begin{figure*}
\includegraphics[width=\widefigwidth]{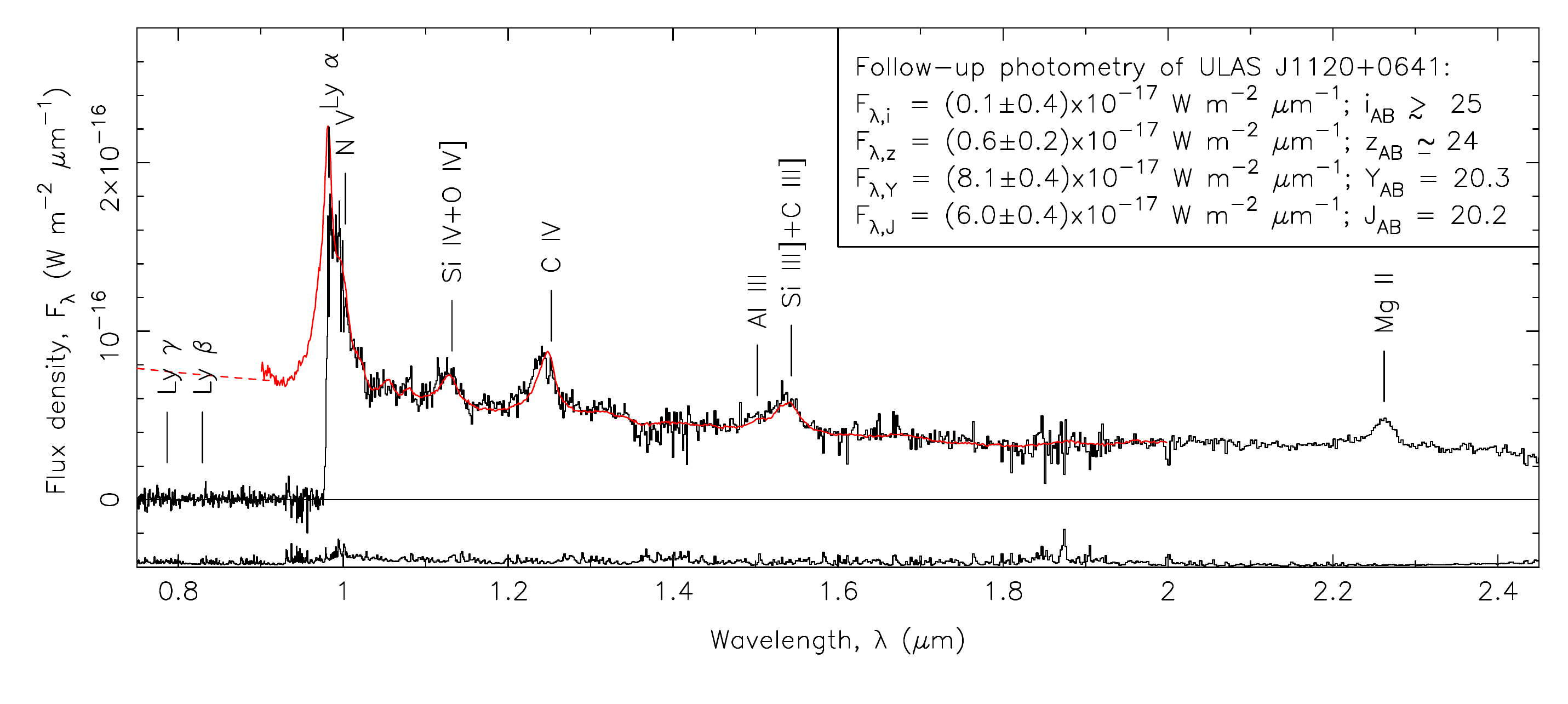}
\caption{$|$ 
{\bf{Spectrum of \newqso\ compared to a composite spectrum
  derived from lower-redshift quasars.}}
Blueward of $1.005 \unit{\micron}$ the spectrum was obtained with
the FORS2 on the Very Large Telescope (VLT) Antu using 
a $1\farcs0$ wide longslit and
the 600z holographic grism, which has a resolution of 1390;
the resultant dispersion was
$1.6 \times 10^{-4} \unit{\micron}$ per pixel
and the spatial scale was $0\farcs25$ per pixel.
The full FORS2 spectrum covers the wavelength range
$0.75 \unit{\micron} \leq \lambda \leq 1.03 \unit{\micron}$.
Redward of $1.005 \unit{\micron}$ the data were obtained using
the GNIRS on the Gemini North Telecope.
The GNIRS observations were made in cross-dispersed mode
using a $32 \unit{lines}$ per mm grating
and the short camera with a pixel scale of $0\farcs15$ per pixel;
with a $1\farcs0$ slit this provided a resolution of 500.
The full GNIRS spectrum covers the wavelength range
$0.90 \unit{\micron} \leq \lambda \leq 2.48 \unit{\micron}$.
The data are binned by a factor of four and are shown in black;
the 1 $\sigma$ error spectrum is shown below the observed
spectrum.
The wavelengths of
common emission lines, redshifted by $\redshift = \seven$,
are also indicated.
The solid red curve shows a
composite spectrum constructed by averaging
the spectra of 169 SDSS quasars
in the redshift interval $2.3 \la \redshift\ \la 2.6$
that exhibit large \civ\ emission line blueshifts.
Absorption lines in the SDSS spectra were masked in forming the composite.
The composite is a strikingly good fit to the spectral shape
of \newqso\
and most of its emission lines,
although it was not possible to match the extreme \civ\ blueshift.
The \lya\ and \civ\ equivalent widths of the SDSS quasars
are strongly correlated; the fact that the equivalent width of
\civ\ from the composite spectrum is similar to that of
\newqso\
implies that the \lya\ line is also correctly modelled.
The dashed red curve shows the power-law
($F_\lambda \propto \lambda^{-0.5}$) used to
estimate the quasar's ionizing flux
and the follow-up photometry of \newqso\ is also listed.}
\label{figure:spectrum}
\end{figure*}

Further spectroscopic observations of \newqso\ were made
using the 
FOcal Reducer/low dispersion Spectrograph 2
(FORS2) on the Very Large Telescope (VLT) Antu 
and 
the Gemini Near-Infrared Spectrograph (GNIRS)
on the Gemini North Telecope
and the results combined into the spectrum shown in \figspectrum.
The spectrum of \newqso\ is 
similar to those of lower redshift quasars of
comparable luminosity, and comparison to a rest-frame template 
spectrum\cite{Hewett_Wild:2010}
over the wavelength range including the strong 
\siiii$+$\ciii\
and 
\mgii\
emission features gives an accurate systemic redshift of 
$\redshift = \sevenerr$.
The most unusual feature of the spectrum is
the
$2800 \pm 250 \unit{km} \unit{s}^{-1}$
blueshift of the \civ\ emission line,
which is greater than that seen in $99.9 \unit{\%}$ of redshift 
$\redshift \ga 2$ quasars\cite{Richards_etal:2011}.
There is associated absorption 
(visible through the \nv\ doublet 
at $\lambda = 0.999 \unit{\micron}$ and the \civ\ doublet at 
$\lambda = 1.249 \unit{\micron}$), 
indicating the presence of 
material in front of the quasar
flowing out at $1100 \pm 200 \unit{km} \unit{s}^{-1}$.
There is also a narrow absorption line at the \lya\ emission wavelength
that is consistent with a cloud of \hi\ close to the quasar.
If \newqso\ is not significantly magnified by gravitational lensing,
the GNIRS spectrum gives an absolute magnitude 
(measured at $0.1450 \unit{\micron}$ in the rest-frame) 
of $\mabs = -26.6 \pm 0.1$ and,
applying a fiducial bolometric correction\cite{Willott_etal:2010b} of 4.4,
a total luminosity of $L = (6.3 \pm 0.6) \times 10^{13} L_\sun$.
\newqso\ has not been
detected at radio wavelengths,
with a measured flux

\affiliations{}

\noindent
of
$F_\nu = -0.08 \pm 0.13 \unit{mJy}$ 
in the
Faint Images of the Radio Sky at Twenty-Centimeters 
(FIRST) survey\cite{Becker_etal:1995}.
Assuming an
unabsorbed continuum blueward 
of \lya\ of the form
$L_\lambda \propto \lambda^{-0.5}$ 
(as appropriate for a radio-quiet quasar\cite{Telfer_etal:2002})
implies \newqso\ was emitting
ionizing photons 
at a rate of $\Gamma_\ion = 1.3 \times 10^{57} \unit{s}^{-1}$.


Quasars are believed to be powered by accretion onto 
their central black holes.
The black hole's mass can be estimated from the 
quasar's luminosity and its \mgii\ line 
width\cite{Vestergaard_Osmer:2009}.
\newqso\ has 
$L_\lambda = (1.3 \pm 0.1) \times 10^{40} \unit{W} \unit{\micron}^{-1}$ 
at a rest-frame wavelength of $\lambda = 0.3 \unit{\micron}$
and the \mgii\ line 
has a full width at half-maximum of
$3800 \pm 200 \unit{km} \unit{s}^{-1}$, 
implying 
$\mbh = (2.0^{+1.5}_{-0.7}) \times 10^9 M_\sun$
(where the uncertainty is dominated by the empirical scatter in the 
scaling relationship).
The Eddington luminosity for \newqso\ is hence 
$L_{\rm{Edd}} = (5.3^{+3.9}_{-1.8}) \times 10^{13} L_\sun$,
which is comparable to the above bolometric luminosity 
and implies an Eddington ratio of $\lambda_{\rm{Edd}} = 1.2^{+0.6}_{-0.5}$.
Assuming Eddington-limited accretion with an efficiency of
$\epsilon \simeq 0.1$,
a black hole's mass would grow as\cite{Volonteri_Rees:2006}
$\mbh \propto \exp(t / (0.04 \unit{Gyr}))$;
this implies that
all other known high-redshift quasars
(for example, SDSS~J1148$+$5251 at $\redshift = 6.42$,
with an estimated\cite{Willott_etal:2003} black hole mass of
$\mbh \simeq 3 \times 10^9 M_\sun$)
would have had $\mbh \la 5 \times 10^8 M_\sun$
at $0.77 \unit{Gyr}$ after the Big Bang.
The
existence of $\simm 10^9 M_\sun$ black holes at $\redshift \simeq 6$
already placed strong limits on the possible models of black hole
seed formation, accretion mechanisms and merger 
histories\cite{Volonteri_Rees:2006,Haiman:2010};
the discovery that 
a $2 \times 10^9 M_\sun$ black hole
existed just $0.77 \unit{Gyr}$ after the Big Bang 
makes these restrictions even more severe.


\begin{figure*}
\includegraphics[width=\widefigwidth]{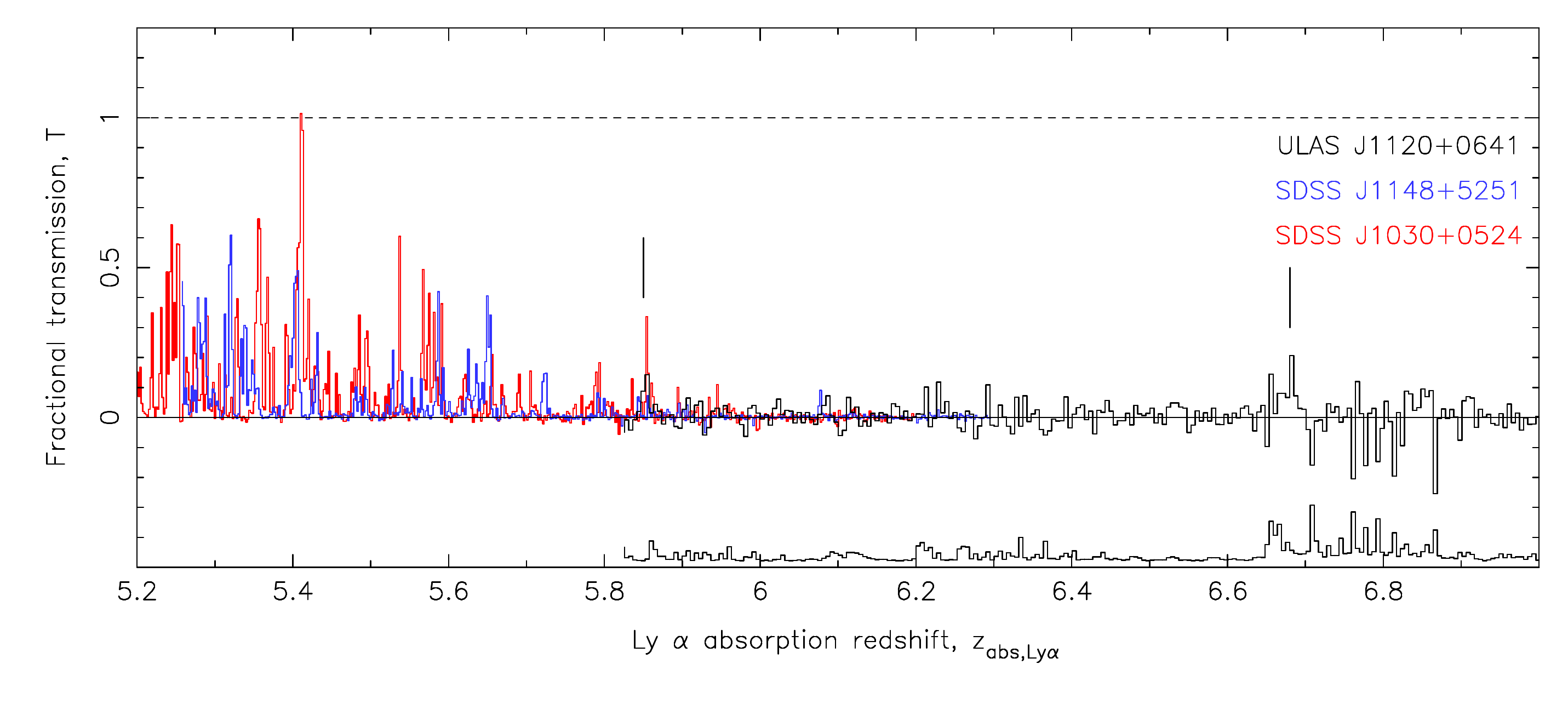}
\caption{$|$ 
{\bf{The observed Ly\,\boldmath{$\alpha$} absorption measured towards 
\newqso\ and two lower-redshift quasars.}}
For \newqso,
SDSS~J1148+5251\protect\cite{Fan_etal:2003} ($\redshift = 6.42$)
and
SDSS~J1030+0524\protect\cite{Fan_etal:2001} ($\redshift = 6.31$)
the transmission was calculated by dividing the
measured spectrum by a power-law continuum
of the form $F_\lambda \propto \lambda^{-0.5}$;
the \lya\ absorption redshift is given by
$\redshift_{\abs,\lya} = \lambda / \lambda_\lya - 1$,
where $\lambda_\lya = 0.12157 \unit{\micron}$.
The three transmission curves are shown as far as
the edges of the quasars' near zones, just blueward of \lya;
the transmission towards \newqso\ is only shown
redward of its \lyb\ emission line,
which corresponds to $\redshift_{\abs,\lya} = 5.821$.
The transmission spectrum of \newqso\ is binned by a factor of four
and the 1 $\sigma$
uncertainty in each pixel is shown below the data.
The measurements\protect\cite{White_etal:2003} of the other two quasars
have a signal-to-noise ratio sufficiently high that the errors can be
ignored.
Comparing the three transmission curves reveals a clear trend:
the numerous transmission spikes at $\redshift \simeq 5.5$
give way to increasingly long Gunn--Peterson\protect\cite{Gunn_Peterson:1965}
troughs at $\redshift \simeq 6$ and,
finally, to almost complete absorption
beyond $\redshift \simeq 6.3$.
The most significant transmission
spikes identified towards \newqso\ are
indicated by the vertical lines:
the feature at
$\redshift_{\abs,\lya} = 5.85$
is detected at $\sim 5.8 \sigma$ over three pixels;
the feature at
$\redshift_{\abs,\lya} = 6.68$
is detected at
$\sim 3.8 \sigma$ over three pixels.
Although the spectrum of \newqso\ has a higher noise level than those
of the two lower-redshift quasars,
transmission spikes of the strength seen at $\redshift_{\abs,\lya} \la 5.8$
would have been clearly detected at $\redshift_{\abs,\lya} \ga 6$
towards \newqso\ if present.}
\label{figure:igm}
\end{figure*}

Aside from its existence, 
the most striking aspect of \newqso\ is 
the almost complete lack of observed flux blueward of 
its \lya\ emission line,
which can be attributed to absorption
by \hi\ along the line of sight.
The transmission, $T$, was quantified by dividing 
the observed spectrum of \newqso\ by 
the power-law shown in \figspectrum.
Converting from observed wavelength to \lya\ absorption redshift 
yields the transmission spectrum shown in \figigm.
The effective optical depth,
defined in the absence of noise as $\taueff \equiv -\ln(T)$,
was measured in redshift bins of width
$\Delta \redshift_{\abs,\lya} = 0.15$ 
in the range $5.9 \la \redshift_{\abs,\lya} \la 7.1$.
In all eight bins the $2 \sigma$ lower limit
is $\taueff > 5$.
The overall implication is that
the neutral hydrogren density at redshifts of 
$\redshift \ga 6.5$ 
was so high that it cannot be probed effectively 
using continuum \lya\ absorption measurements.


\begin{figure}
\includegraphics[width=\figwidth]{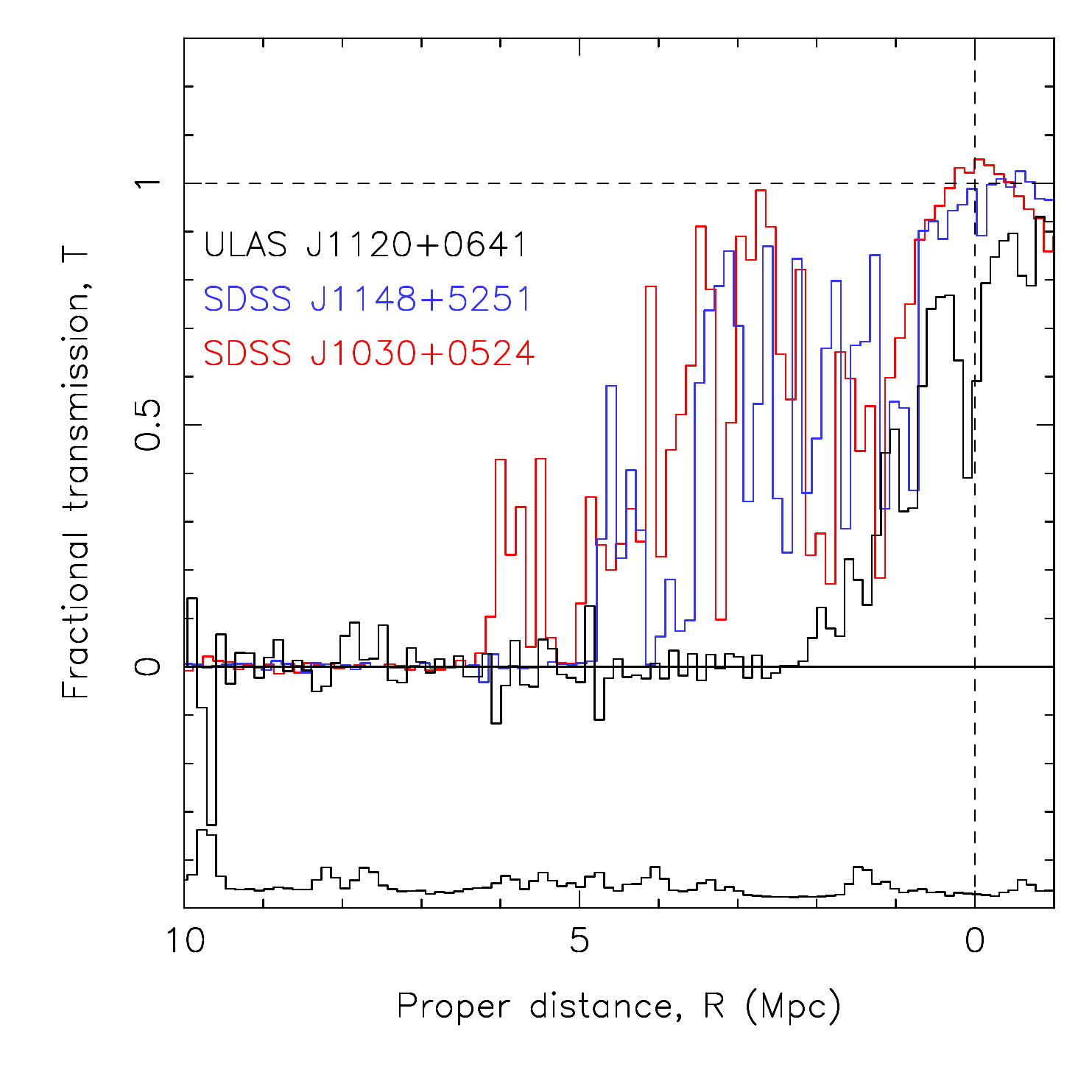}
\caption{$|$ 
{\bf{The inferred Ly\,\boldmath{$\alpha$} near zone transmission profile of
 \newqso\ compared to those of two lower-redshift quasars.}}
The near zone transmission profile of \newqso\
was estimated by dividing the observed
spectrum by the composite spectrum shown in \figspectrum.
The transmission profiles towards the two SDSS quasars 
were
estimated by dividing their measured\protect\cite{White_etal:2003} spectra
by parameterised fits based on the unabsorbed spectra of
lower-redshift quasars.
The transmission profile of \newqso\
is strikingly different from those of the two SDSS quasars,
with a much smaller observed near zone radius,
$\rnz$, as well as a distinct shape:
whereas the profiles of SDSS~J1148$+$5251
and SDSS~J1030$+$0524 have approximately Gaussian envelopes
out to a sharp cut-off, the profile of \newqso\
is much smoother and also shows absorption redward of \lya.
The 1 $\sigma$ deviation error spectrum for \newqso\ is
shown below the data.}
\label{figure:proxzone}
\end{figure}

The inability of \lya\ forest absorption measurements to probe
high optical depths is generic,
but quasars differ from other high-redshift sources
in that they have a strong effect 
on the intergalactic medium in their vicinity,
ionizing megaparsec-scale near zones around them.
Ultraviolet photons can propagate freely through these
ionized regions, resulting in significant transmission just 
blueward of the \lya\ emission wavelength.
The scale of the near zone can be
characterised by\cite{Fan_etal:2006b} $\rnz$,
the (proper) radius at which the
measured transmission drops to $T = 0.1$,
and then corrected to $\rnzcorr = 10^{0.4\,(27 + \mabs) / 3} \, \rnz$
to compare quasars of different luminosities.
The near zone transmission profile of \newqso,
shown in \figproxzone,
implies that 
$\rnz = 1.9 \pm 0.1 \unit{Mpc}$
and
$\rnzcorr = 2.1 \pm 0.1 \unit{Mpc}$.
This is considerably smaller than 
the near zones of other comparably luminous high-redshift quasars,
which have been measured\cite{Carilli_etal:2010} to have
$\rnzcorr = (7.4 - 8.0 (\redshift - 6)) \unit{Mpc}$
on average.
The considerable scatter about this trend notwithstanding, 
these observations of
\newqso\ confirm that 
the observed decrease in $\rnzcorr$ with redshift
continues at least to $\redshift \simeq 7.1$.


The observed transmission cut-offs of 
$\redshift \simeq 6$ quasars have been
identified with their
advancing ionization fronts,
which grow as\cite{Haiman:2002,Bolton_Haehnelt:2007}
$\rnzcorr \propto \tq^{1/3} (1 + \redshift)^{-1} \Delta^{-1/3} \nf^{-1/3}$,
where $\tq$ is the quasar age and 
$\Delta$ is the local baryon density relative to the cosmic mean.
Assuming 
a fiducial age of $\tq \simeq 0.01 \unit{Gyr}$
has led to the 
claim\cite{Wyithe_etal:2005}
that $\nf \ga 0.6$
around several redshift $6.0 \la \redshift \la 6.4$ quasars.
Given that the above $\rnzcorr$-$\redshift$ fit
gives an average value of $\rnzcorr = 5.8 \unit{Mpc}$ at 
$\redshift = 6.2$,
the measured near zone radius of \newqso\ 
then implies that the neutral fraction was a
factor of $\simm 15$ higher 
at $\redshift \simeq 7.1$ than it was at $\redshift \simeq 6.2$.
The fundamental limit that $\nf \leq 1$, 
makes it difficult to reconcile 
the small observed near zone of \newqso\ with 
a significantly neutral Universe at $\redshift \simeq 6$.
It is possible that \newqso\ is seen very early in its luminous phase 
or that it formed in an unusually dense region,
but the most straightforward conclusion is that 
observed near zone sizes of $\redshift \simeq 6$ quasars 
do not correspond to their ionization fronts\cite{Bolton_Haehnelt:2007}.


\begin{figure}
\includegraphics[width=\figwidth]{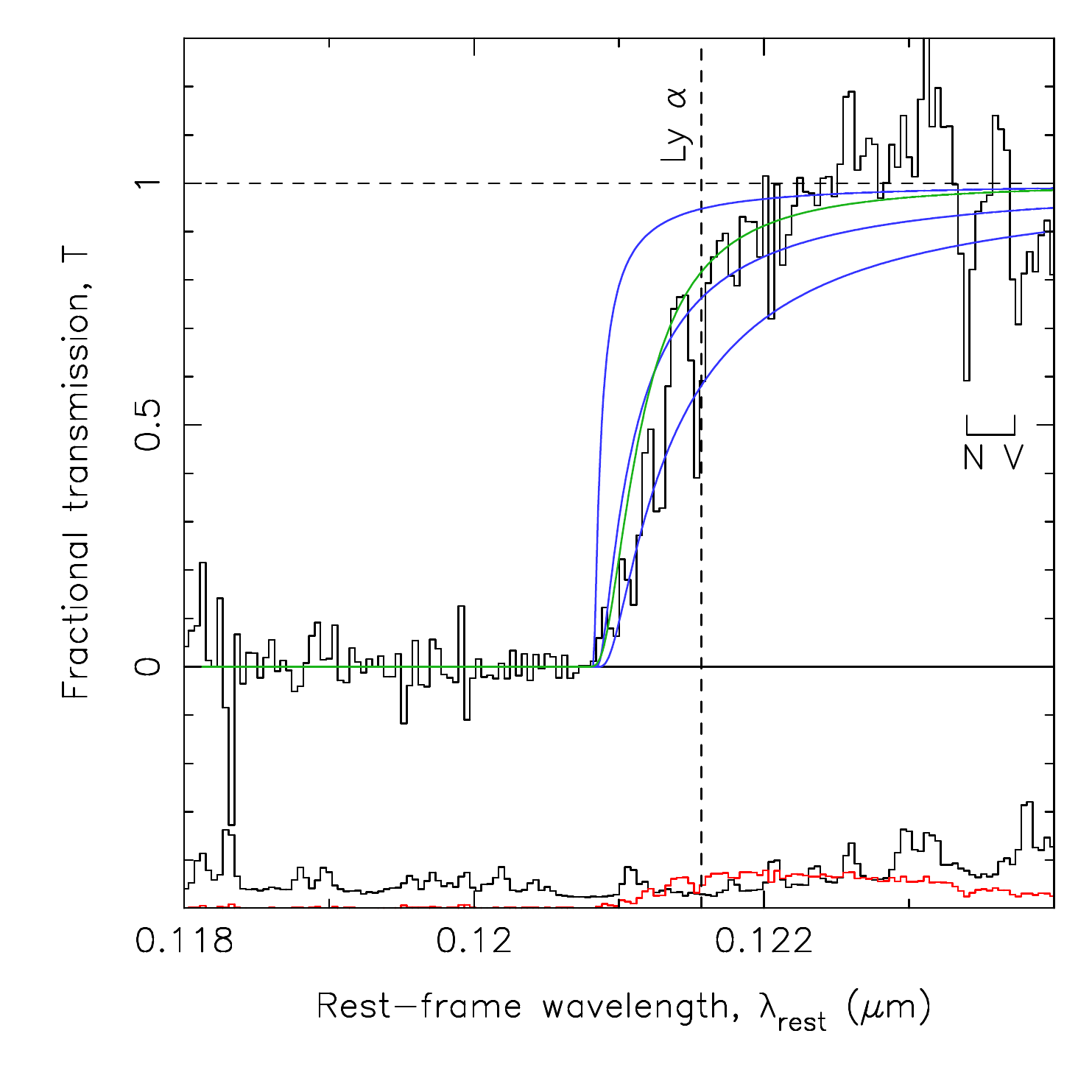}
\caption{$|$ 
{\bf{Rest-frame transmission profile of \newqso, in the region of the
  Ly\,\boldmath{$\alpha$} alpha emission line, compared to several
  damping profiles.}}
The transmission profile of \newqso,
obtained by dividing the spectrum by the
SDSS composite shown in \figspectrum, is shown in black.
The random error spectrum is plotted below the data,
also in black.
The positive residuals near $0.1230 \unit{\micron}$ in the transmission
profile suggest
that the \lya\ emission line of \newqso\ is actually stronger than average,
in which case the absorption would be greater than illustrated.
The dispersion
in the \lya\ equivalent width 
at fixed \civ\ equivalent width
of $13 \unit{\%}$ 
quantifies the uncertainty in the \lya\ strength;
this systematic uncertainty in the
transmission profile is shown in red.
The blue curves show the \lya\ damping wing of the IGM for neutral
fractions of (from top to bottom) $\nf = 0.1$, $\nf = 0.5$ and $\nf = 1.0$,
assuming a sharp ionization front $2.2 \unit{Mpc}$
in front of the quasar.
The green curve shows the absorption profile of a damped \lya\ absorber of
column density $N_{\rm{H\,I}} = 4 \times 10^{20} \unit{cm}^{-2}$
located $2.6 \unit{Mpc}$ in front of the quasar.
These curves assume that the ionized zone itself is completely transparent;
a more realistic model
of the \hi\ distribution around the quasar
might be sufficient to discriminate between these
two models\protect\cite{Mesinger_Haiman:2004,Bolton_Haehnelt:2007}.
The wavelength of the \lya\
transition is shown as a dashed line;
also marked is the
\nv\ doublet of the associated absorber referred to in the text.}
\label{figure:wing}
\end{figure}

An alternative explanation for the near zones of 
the $\redshift \simeq 6$ quasars
is that their transmission profiles 
are determined primarily by the residual \hi\ 
inside their ionized zones\cite{Mesinger_Haiman:2004,Bolton_Haehnelt:2007}.
If the \hi\ and \hii\ are in equilibrium with the ionizing
radiation from the quasar then 
the neutral fraction would increase with radius
as $\nf \propto R^2$ out to the ionization front.
The resultant transmission profile would have an
approximately Gaussian envelope,
with 
$\rnz$ being the radius at which\cite{Bolton_Haehnelt:2007}
$\nf \simeq 10^{-4}$, and not the ionization front itself.
The envelopes of the measured profiles of the two 
$\redshift \simeq 6.3$ quasars shown in \figproxzone\
are consistent with this Gaussian model,
although both have sharp cut-offs as well,
which could be due to 
Lyman limit systems along the 
line of sight\cite{Calverley_etal:2011}.

In contrast,
the measured transmission profile of \newqso,
shown in \figproxzone,
is qualitatively different from those of 
the lower redshift quasars,
exhibiting a smooth envelope
and significant absorption redward of the \lya\ wavelength.
The profile has the character of
a \lya\ damping wing,
which would indicate that the 
intergalactic medium in front of \newqso\ was substantially neutral.
It is also possible that the absorption is the result of 
an intervening high column density 
($N_{\rm{H\,I}} \ga 10^{20} \unit{cm}^{-2}$) 
damped \lya\ system\cite{Miralda-Escude:1998},
although absorbers of such strength are rare.
Both models are compared to 
the observed transmission profile 
of \newqso\ in \figwing.
Assuming the absorption is the result of the IGM damping wing,
the shape and width of the transmission profile 
require $\nf > 0.1$, 
but are inconsistent with $\nf \simeq 1$,
at $\redshift \simeq 7.1$.
These limits will be improved by 
more detailed modelling,
in particular accounting for the distribution of \hi\ within the near 
zone\cite{Mesinger_Haiman:2004,Bolton_Haehnelt:2007},
and deeper spectroscopic observations of \newqso.
Given the likely variation in the ionization
history between different lines of sight,
it will be important to find more sources in the 
epoch of reionization.
However,
there are only expected\cite{Willott_etal:2010} to be $\simm 10^2$
bright quasars with $\redshift \ga 7$ over the whole sky,
\newqso\ will remain a vital probe of the early Universe for some time.


\vspace*{5mm}
\noindent
{\sf{\textbf{Received 11 March 2011; accepted 28 April 2011.}}}
\vspace*{-12mm}



\bibliographystyle{nature_djm.bst}
\bibliography{references}



\balance

\vspace*{2mm}
\noindent
{\sf{\textbf{Acknowledgements}}
M.P.\ acknowledges support from the University of London's Perren Fund.
P.C.H.\ and R.G.McM.\ acknowledge support from the STFC-funded
Galaxy Formation and Evolution programme at the Institute of Astronomy.
X.\ Fan and R.\ White kindly supplied spectra of the SDSS quasars.
M.\ Haehnelt and J.\ Bolton
provided valuable insights into quasar near zone physics.
The careful reading of this paper by two referees helped clarify several
important issues.
The staffs of the Joint Astronomy Centre,
the Cambridge Astronomical Survey Unit
and
the Wide-Field Astronomy Unit, Edinburgh,
all made vital contributions to the UKIDSS project.
The support staff at the Gemini North Telescope,
particularly K.\ Roth, provided invaluable assistance
with the Gemini observations.
This work is based in part on data obtained from the
UKIDSS,
SDSS,
the Liverpool Telescope,
the Isaac Newton Telescope,
the Gemini Observatory and
the European Southern Observatory.}

\vspace*{2mm}
\noindent
{\sf{\textbf{Author Contributions}}
D.J.M., S.J.W., M.P., B.P.V., P.C.H., R.G.McM.\ and C.S.\
identified \newqso\ and obtained the follow-up observations.
S.J.W., P.C.H., D.J.M., T.T., B.P.V., R.G.McM.\ and M.P.\
analysed the follow-up observations and interpreted the results.
A.A., S.D., E.G.-S., N.C.H., P.H., M.J.I.\ and A.L.\
obtained, analysed and disseminated the UKIDSS data.
E.K.\ and H.J.A.R.\ obtained the FORS2 spectrum of \newqso.
D.J.M.\ and S.J.W.\
wrote the manuscript, into which all other authors had input.}

\vspace*{2mm}
\noindent
{\sf{\textbf{Author Information}}
Reprints and permission information is available at 
www.nature.com/reprints.
The authors declare no competing financial interests.
Readers are welcome to comment on the online version of this article
at www.nature.com/nature.
Correspondence and requests for materials
should be addressed to D.J.M.\ (mortlock@ic.ac.uk).}


\label{lastpage}
\end{document}